\documentclass[10pt]{article}
\usepackage{graphicx}

\def\Title#1{\begin{center} {\Large #1 } \end{center}}
\def\Author#1{\begin{center}{ \sc #1} \end{center}}
\def\Address#1{\begin{center}{ \it #1} \end{center}}

\newcommand\pubblock{\rightline{\begin{tabular}{l} Proceedings of the Second Annual LHCP\\ \pubnumber\\
         \pubdate  \end{tabular}}}

\newenvironment{Abstract}{\begin{quotation} \begin{center} 
             \large ABSTRACT \end{center}\bigskip 
      \begin{center}\begin{large}}{\end{large}\end{center} \end{quotation}}

\newenvironment{Presented}{\begin{quotation} \begin{center} 
             PRESENTED AT\end{center}\bigskip 
      \begin{center}\begin{large}}{\end{large}\end{center} \end{quotation}}

\def\beq{\begin{equation}}
\def\eeq#1{\label{#1}\end{equation}}
\def\eeqn{\end{equation}}

\def\beqa{\begin{eqnarray}}
\def\eeqa#1{\label{#1}\end{eqnarray}}
\def\eeqan{\end{eqnarray}}

\let\bar=\overbar

\def\Dslash{\not{\hbox{\kern-4pt $D$}}}
\def\dslash{\not{\hbox{\kern-2pt $\del$}}}

\def\msb{{\bar{\ssstyle M \kern -1pt S}}}

\usepackage{color}

\newcommand\pubnumber{}

\newcommand\pubdate{\today}

\def\affiliation{
On behalf of the CMS Experiment, \\
INFN and University of Torino, Italy}

\begin{document}

\large
\begin{titlepage}

\pubblock

\vfill
\Title{Recent Results on the Higgs Boson Properties in the $H\to ZZ \to 4\ell$ decay channel at CMS}
\vfill

\Author{ LINDA FINCO  }
\Address{\affiliation}
\vfill
\begin{Abstract}

The latest results on the measurement of the properties of the new boson with mass around 125 GeV are reported. 
The analysis uses pp collision data recorded by the CMS detector at the LHC, corresponding to integrated 
luminosities of 5.1 $fb^{-1}$ at $\sqrt{s}$ = 7 TeV and 19.6 $fb^{-1}$ at $\sqrt{s}$ = 8 TeV. The boson is
observed in the $H \to ZZ \to 4\ell$ channel ($\ell = e, \mu$) and its mass is measured, giving the most precise
result ever achieved. Moreover, the first experimental constraint on Higgs total width using $H \to ZZ \to 4\ell$ events is 
presented, setting an upper limit of 33 MeV at 95\% confidence level (42 MeV expected). The spin-parity of the
boson is studied and the pure scalar hypothesis is found to be consistent with the observation, when compared to 
the other spin-parity hypotheses. No other significant Standard Model Higgs-like excess is found in the search and 
upper limits at 95\% condence level exclude the range 129.5 -832.0 GeV. 

\end{Abstract}
\vfill

\begin{Presented}
The Second Annual Conference\\
 on Large Hadron Collider Physics \\
Columbia University, New York, U.S.A \\ 
June 2-7, 2014
\end{Presented}
\vfill
\end{titlepage}
\def\thefootnote{\fnsymbol{footnote}}
\setcounter{footnote}{0}
%

\normalsize 


\section{Introduction}

The latest results on the measurement of the properties of the new boson with mass around 
125 GeV \cite{Atlas:Higgs} \cite{CMS:Higgs} are measured in the $H \to ZZ \to 4\ell$ decay channel 
($\ell = e, \mu$). 
The analysis uses pp collision data recorded by the CMS detector \cite{CMS:Detector} at the LHC, corresponding to integrated 
luminosities of 5.1 $fb^{-1}$ at $\sqrt{s}$ = 7 TeV and 19.6 $fb^{-1}$ at $\sqrt{s}$ = 8 TeV.

\section{Analysis Strategy}

The $H\to ZZ \to 4\ell$ analysis \cite{CMS:Prop} is based on the reconstruction, identification
and isolation of leptons. Each signal event consists of two pairs 
of same-flavor and opposite-charge leptons in the final state, isolated and with high 
transverse momentum, and it is compatible with a ZZ system, where one or both Z bosons can be off-shell.
The sources of background for the $H\to ZZ \to 4\ell$ channel are the
irreducible four-lepton contribution from direct $ZZ$ (or $Z\gamma ^*$) production, very 
similar to the signal, the reducible background arising from $Zb\bar{b}$ and 
$t\bar{t}\to 4\ell$ decays and the instrumental contribution due to a misidentification
of the leptons.\\
\\
In order to separate signal from background events, a kinematic discriminant is defined 
($\mathcal{D}_{bkg}^{kin}$), depending on
the five production and decay angles 
and the $Z$ boson masses. These variables
fully describe the event topology and have a high discriminating power. 
The $\mathcal{D}_{bkg}^{kin}$ discriminant \cite{CMS:MELA} is defined as
$$\mathcal{D}_{bkg}^{kin} = \frac{\mathcal{P}_{sig}^{kin}}{\mathcal{P}_{sig}^{kin} +\mathcal{P}_{bkg}^{kin}},$$
where $\mathcal{P}^{kin}_{sig\ (bkg)}$ is the probability for an event with a given topology (angles and masses)
to come from a signal (background) process.

\section{Significance and Signal Strenght}
The minimum of the local p-value is reached at $m_{4\ell}$ = 125.7 GeV (Fig. \ref{fig:Pmass}, left) and it corresponds to a local significance 
of 6.8 (for an expectation of 6.7). This is the only significant excess in the range $m_H <$ 1 TeV.\\
The parameter that describes the magnitude of the 
Higgs signal is the signal strength modifier, defined as 
the ratio of the observed cross section and the cross 
section predicted by the SM ($\mu = \sigma _{obs}/\sigma _{SM}$). The measured value of $\mu$ obtained at 
the best fit mass ($m_H$ = 125.6 GeV) is: 
$$\mu = 0.93 ^{+0.26}_{-0.23}(stat.) ^{+0.13}_{-0.09}(syst.).$$

\section{Mass Measurement}
\begin{figure}[htb]
\centering
\includegraphics[height=2in]{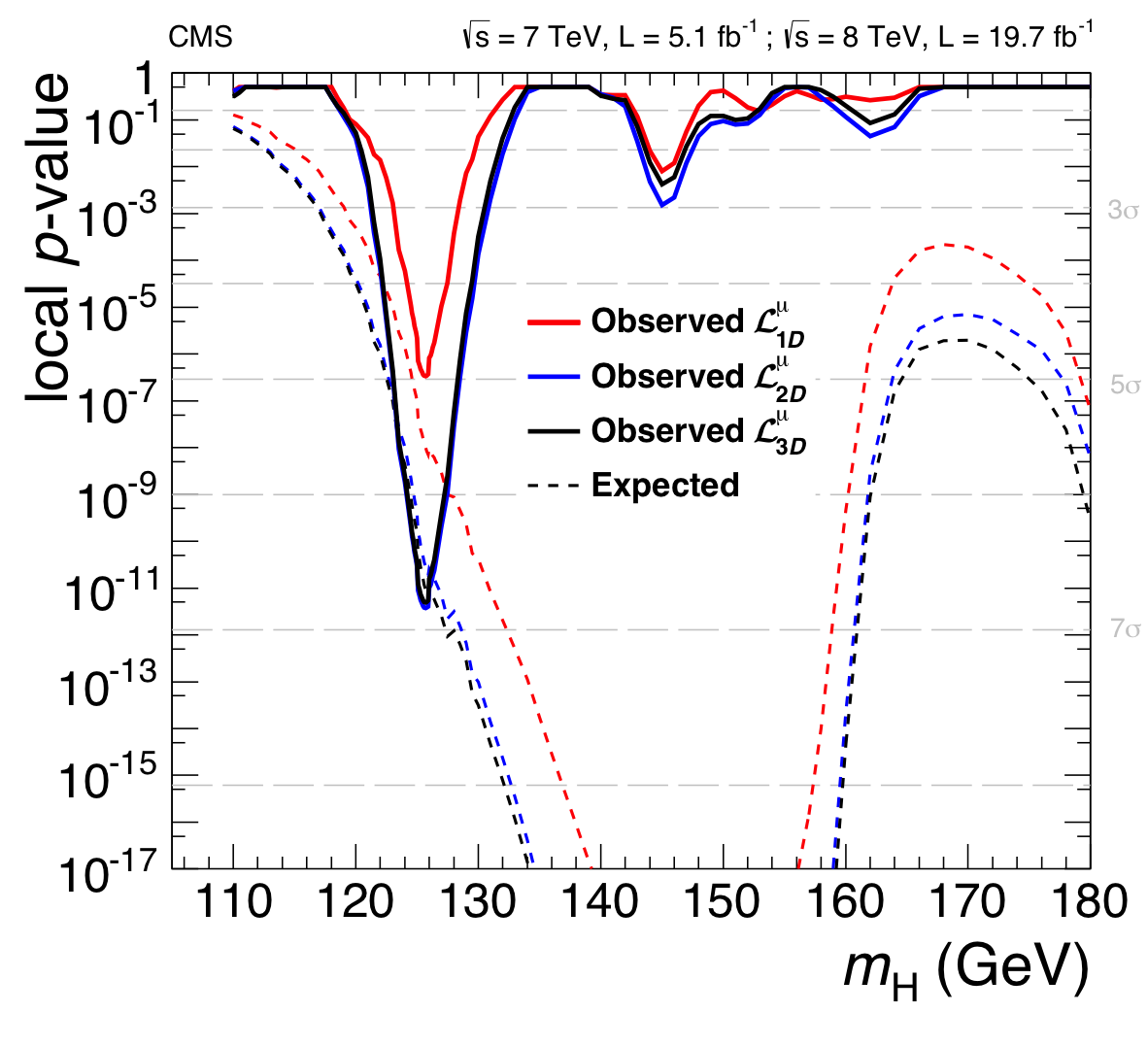}
\includegraphics[height=2in]{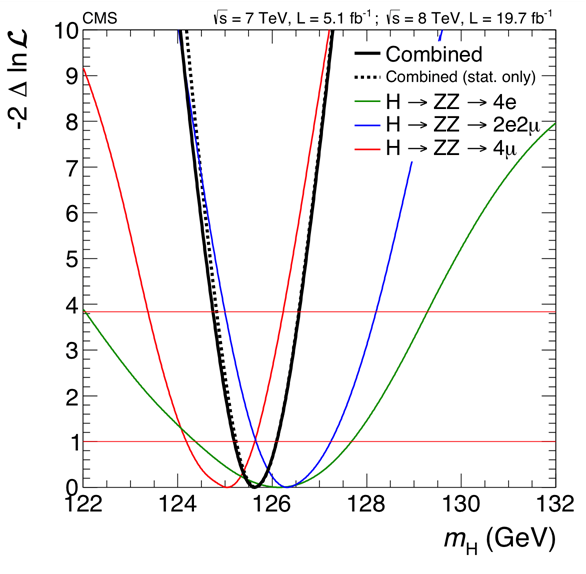}
\caption{Significance of the local excess with respect to the SM background expectation as a
function of the Higgs boson mass in the low mass range (left): red is 1D model ($m_{4\ell}$), blue is 2D model
($m_{4\ell}$, $\mathcal{D}_{bkg}^{kin}$) and black is 3D model ($m_{4\ell}$, $\mathcal{D}_{bkg}^{kin}$, $\mathcal{D}_{jet}$ or $p_T^{4\ell}$). 
Likelihood scan as a function of mass obtained from the 3D test statistics for the different final states and their combination (right).}
\label{fig:Pmass}
\end{figure}

The mass measurement is performed with a three-dimensional fit using for each event the four-lepton invariant
mass ($m_{4\ell}$), the associated per-event mass error ($\mathcal{D}_m$) and the kinematic 
discriminant ($\mathcal{D}_{bkg}^{kin}$). Per-event errors are calculated from the individual lepton momentum 
errors and including them in the fit allows to gain 8\% improvement in the Higgs boson mass measurement 
uncertainty. The fit procedure gives $m_H = 125.6 \pm 0.4 (stat.) \pm 0.2 (syst.)$ GeV (see Fig. \ref{fig:Pmass}, right). 

\section{Spin-Parity Measurement}
\begin{figure}[htb]
\centering
\includegraphics[height=2in]{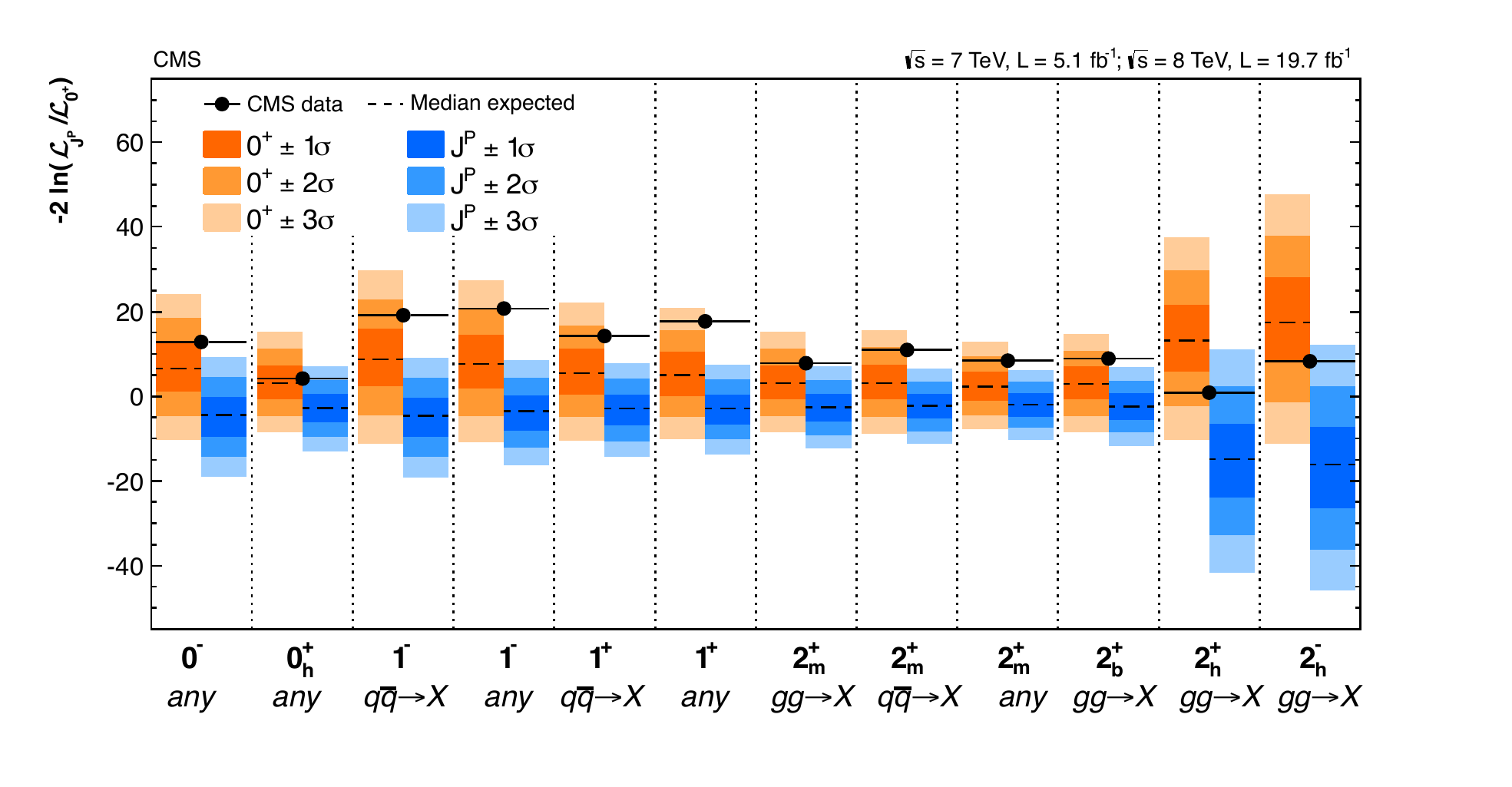}
\caption{Summary of the expected and observed values for the test-statistic $q$ distributions for the twelve alternative hypotheses
tested with respect to the SM Higgs boson.}
\label{fig:spin_ZZ}
\end{figure}

In order to determine the spin and the parity of the new boson, a methodology with kinematic discriminants is 
used. Two discriminants are defined, in order to separate SM Higgs from background events ($\mathcal{D}_{bkg}$)
and to discriminate an alternative hypothesis from the SM Higgs ($\mathcal{D}_{J^P}$). The different spin-parity 
hypotheses are thus tested using the two-dimensional likelihood 
$\mathcal{L}_{2D} =\mathcal{L}_{2D}(\mathcal{D}_{J^P},\mathcal{D}_{bkg})$.\\
The distribution of the test statistic $q = -2ln(\mathcal{L}_{J^P}/\mathcal{L}_{SM})$ is
determined and it is examined with generated samples for $m_H$ = 125.6 GeV.\\
A confident levels (CLs) criterion is defined as the ratio of the probabilities to observe, under the $J^P$ and 
$0^+$ hypotheses, a value of the test statistic $q$ equal or larger than the one in the
data. The data disfavor the alternative hypotheses $J^P$ with a CLs value in the range 0.001 -10\% (see Fig. \ref{fig:spin_ZZ}).

\section{Width Measurement}
\begin{figure}[htb]
\centering
\includegraphics[height=2in]{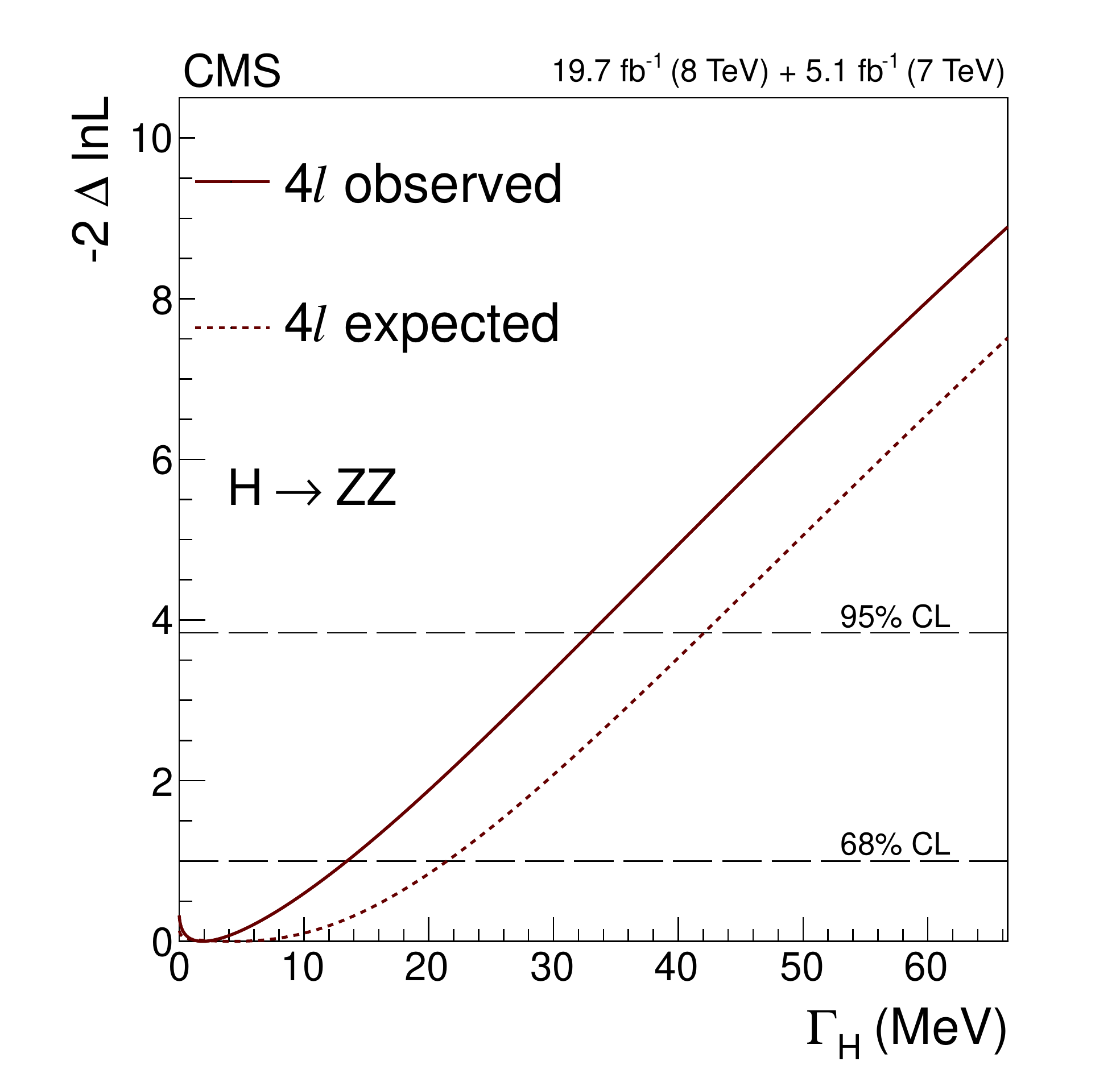}
\caption{Likelihood scan of the $\Gamma _H$ in the 4$\ell$ final state.}
\label{fig:width}
\end{figure}

At $m_H$ = 125.6 GeV, the Standard Model predicts a Higgs boson decay width ($\Gamma _H$) of 4.15 MeV. A 
direct measurement at the resonance peak is thus strongly limited by experimental resolution, but it is
possible to constrain the Higgs boson width using its off-shell production and decay away from 
the resonance  \cite{CMS:Width}. Indeed, the integrated cross 
sections in the resonant and off-shell regions are
$$\sigma ^{on-shell}_{gg \to H \to ZZ} \sim \frac{g^{2}_{ggH} g^{2}_{HZZ}}{m_H \Gamma _H}, \ \ \ \ 
\sigma ^{off-shell} _{gg \to H \to ZZ} \sim \frac{g^2_{ggH}g^2_{HZZ}}{(2m_Z)^2},$$
where $g_{ggH}$ and $g_{HZZ}$ are the couplings of the Higgs boson to gluons and Z bosons, respectively.
Therefore, the value of $\Gamma _H$ can be extracted by measuring the ratio of the production in the off-shell
and on-shell region, taking into account the destructive interference with continuum $gg \to ZZ$, which is not
negligible at high masses. In order to separate $gg \to ZZ$ events from the $q\bar{q} \to ZZ$ process,
the dominant background of the analysis, a kinematic discriminant is built ($\mathcal{D}_{gg}$).\\
A likelihood function is defined for both the off-shell and the on-shell region, depending on the
total probability distribution functions
$$P^{off-shell}_{tot} = \mu _{ggH}\times (\Gamma _H/ \Gamma _0) \times P ^{gg}_{sig} + \sqrt{\mu _{ggH}\times 
(\Gamma _H/ \Gamma _0)} \times  P ^{gg}_{int} +  P ^{gg}_{bkg} + P ^{q\bar{q}}_{bkg} + ...$$
$$P^{on-shell}_{tot} = \mu _{ggH}\times P ^{gg}_{sig} + P ^{gg}_{bkg} + P ^{q\bar{q}}_{bkg} + ...$$
and the parameters $\Gamma _H$  and $\mu _{ggH}$ are left unconstrained in the fit. The simultaneous maximum
 likelihood fit leads to an observed (expected) upper limit of $\Gamma _H\ <$ 33 MeV (42 MeV) at 95\% C.L., i.e.
 8.0 (10.1) times the Standard Model prediction (Fig. \ref{fig:width}).



\end{document}